\documentclass[usenatbib,useAMS]{mn2e}
\usepackage{psfig,longtable}
\usepackage{amsmath}

\title[The alignment of debris disks and their host stars] {On the
  alignment of debris disks and their host stars' rotation axis --
  implications for spin-orbit misalignment in exoplanetary systems}

\author[C.\,A.\ Watson et al.]  {C.\,A.\ Watson$^1$\thanks{E-mail:
    c.a.watson@qub.ac.uk}, S.\,P.\ Littlefair$^2$, C.\ Diamond$^1$,
  A.\ Collier Cameron$^3$, \newauthor A.\ Fitzsimmons$^1$,
  E.\ Simpson$^1$, V.\ Moulds$^1$,  and D.\ Pollacco$^1$ \\ $^1$
  Astrophysics Research Centre, Queen's University Belfast, Belfast
  BT7 1NN, UK\\ $^2$ Department of Physics and Astronomy, University
  of Sheffield, Sheffield S3 7RH, UK\\ $^3$ School of Physics and
  Astronomy, University of St Andrews, North Haugh, St Andrews, Fife
  KY19 9SS, UK \\}

\date{\center{\Large Submitted for publication in Letters of the
    Monthly Notices of the Royal Astronomical Society \\
\vspace{.5cm} \today}}

\begin{document}
\maketitle

\begin{abstract}

It has been widely thought that measuring the misalignment angle
between the orbital plane of a transiting exoplanet and the spin of
its host star was a good discriminator between different migration
processes for hot-Jupiters. Specifically, well-aligned hot-Jupiter
systems (as measured by the Rossiter-McLaughlin effect) were thought
to have formed via migration through interaction with a viscous disk,
while misaligned systems were thought to have undergone a more violent
dynamical history. These conclusions were based on the assumption that
the planet-forming disk was well-aligned with the host star. Recent
work by a number of authors has challenged this assumption by
proposing mechanisms that act to drive the star-disk interaction
out of alignment during the pre-main sequence phase. We
have estimated the stellar rotation axis of a sample of stars which
host spatially resolved debris disks. Comparison of our derived
stellar rotation axis inclination angles with the geometrically
measured debris-disk inclinations shows no evidence for a misalignment
between the two.

\end{abstract}


\section{Introduction}
\label{sec:intro}

The discovery of planets beyond the confines of our Solar system has
presented many surprises and continues to challenge our understanding
of planet formation and their subsequent evolution. This is
particularly true in the case of hot-Jupiters, whose short orbital
periods of a few days or less was unexpected -- under the standard
core-accretion theory of planet formation, volatile gas-giants should
form beyond the snow-line (\citealt{pollack96}). It is now widely
accepted that hot-Jupiters did not form in-situ at their current
locations, but that some mechanism caused their inwards migration
towards their parent star.

A number of theories have been postulated to explain planetary
migration.  One possible mechanism for forming short-period gas-giants
is the pumping of initially wide circular orbits to high
eccentricities. This could occur via planet-planet scattering
(\citealt{rasio96}; \citealt{weidenschilling96}), or perturbations
from a distant stellar binary companion (\citealt{eggenberger04}). The
highly eccentric orbit then brings the gas-giant sufficiently close to
the host star that tidal dissipation quickly draws the planet to a
new, smaller orbital separation. In this scenario, the interactions
and scattering involved may lead to large changes in the value of the
orbital inclination. Interactions between the planet and a viscous
disk, on the other-hand, may also drive the planet inwards but is not
thought to perturb the initial orbital inclination.

The close alignment of the rotation and orbital axes in the Solar
system ($\sim$7$^{\circ}$; \citealt{beck05}) is attributed to the
formation of the Sun and planets from a single rotating proto-stellar
disk which was also initially coplanar to the solar-rotation axis. On
the premise that disks and stellar rotation axes are aligned,
Rossiter-McLaughlin (RM) observations of transiting systems
(e.g. \citealt{triaud10} and references therein) have sought to
discriminate between migration caused by planet-disk interactions
(leading presumably to aligned systems), and migrations involving some
violent dynamical history (leading to misaligned systems).  Recent
theoretical work, however, challenges the view that the stellar spin
axis and the disk rotation axis should be aligned. For example, in
numerical simulations of star formation, \cite*{bate10} found that the
rotation axis of the final disk may be heavily governed by the angular
momentum component of the material that was last accreted. While this
can lead to a misalignment between the disk and star, it is only
thought to be significant for light disks which may have insufficient
mass to form  giant planets.

On the other-hand, \cite{Lai2010} present arguments that the observed
star-orbit misalignment could instead result from alterations in the
{\em stellar} spin axis, introduced by the star-disk interaction
during the pre-main-sequence phase (also see \citealt{foucart10}). Lai
et al. consider the well known fact that a magnetic protostar exerts a
warping force on the inner part of the accretion disk
\citep[e.g.][]{Bouvier2007}. Previous authors have assumed that this
results in significant warps to the inner disk, whereas \cite{Lai2010}
argue that viscous processes in the disk itself will smooth these
torques, resulting in a largely unwarped inner disk. Given a flat
disk, the torques arising from the star-disk interaction will act on
the star itself, changing the stellar spin axis on a timescale given
by
\begin{align}
t_{spin} = (1.25 {\rm Myr}) \left(\frac{M_*}{1 M_{\odot}} \right)
\left( \frac{\dot{M}}{10^{-8} M_{\odot} {\rm yr}^{-1}}\right)^{-1}
\notag \\ \times \left( \frac{r_{in}}{4R_*} \right)^{-2}
\frac{\omega_s}{\Omega(r_{in})},
\end{align} 
where $M_*$ and $R_*$ are the mass and radius (in solar units) of the
protostar, respectively, $\dot{M}$ is the accretion rate in solar
masses per year, $r_{in}$ is the inner radius of the accretion disk in
stellar radii, $\omega_s$ is the spin rate of the protostar and
$\Omega(r_{in})$ is the rotation rate of the accretion disk at the
inner disk radius. 
 
However, the mechanism proposed by \cite{Lai2010} may not be effective
in practice, as the timescale for spin evolution, $t_{spin}$, is of
the same order as the disk evolution timescale. Near-infrared
observations of protostars show that the majority of protostellar
disks have dispersed by the age of 5 Myr \citep{Hernandez2008} whilst
observations of the accretion rates onto young stars also show that
the accretion rate declines rapidly with increasing age and decreasing
stellar mass \citep[e.g][]{Sicilia2006}, so the accretion rate on many
protostars may well be below the canonical $10^{-8}$
M$_{\odot}$yr$^{-1}$ assumed by \cite{Lai2010}. In addition, the
results of \cite{Lai2010} rely on the inner disk not being
`significantly warped', however there is good evidence that the inner
disks of some young stars do contain significant disk warps \cite[see
  e.g.][]{Bouvier2007,Muzerolle2009}. 

If correct, these theoretical papers potentially have important
ramifications for our interpretation of the results of RM
observations. Indeed, if the stellar rotation axis can be driven from
coplanarity with the surrounding disk, or vice-versa, then RM
observations would essentially be rendered useless as a tool for
determining the migration mechanism responsible for forming
hot-Jupiter's.  For these reasons, it is important to seek
observational evidence for such processes. In
this paper we present a study of star-disk alignment in debris disk
systems. 

\section{Measuring the star-disk alignment}

For the purposes of this work, we have concentrated on systems with
spatially resolved debris disks. We have also assumed that the
debris-disk plane is representative of the primordial disk and,
likewise, that the presently observed stellar orientation is the same
as the protostars'.  The inclination of the disk to our line-of-sight
can then be measured geometrically by calculating the fore-shortening
of the semi-minor axis of the disk relative to the semi-major axis
(although in reality the models used to determine the disk geometry
are somewhat more complex).

A more indirect approach is needed in order to determine the inclination
angle of the stellar rotation axis, however. To do this we have followed
the method of \cite{watson10} who compiled the stellar rotation inclination
angles for 117 exoplanet host stars, and we refer the reader
to that paper for in-depth details of the methods used, as well as
a discussion on possible sources of systematic errors inherent in
the technique. In summary, it is possible to determine the
inclination angle, $i$, between the rotation axis of a star and the
observers line-of-sight from measurements of the projected equatorial
velocity ($v \sin i$), the stellar rotation period ($P_{rot}$) and the
stellar radius ($R_*$) via the equation

\begin{equation}
\sin i = \frac{P_{rot} \times v \sin i}{2 \pi R_*}.
\label{eqn:sini}
\end{equation}

\noindent The projected equatorial rotation velocity, $v \sin i$,
can be measured using high-resolution spectroscopy, while the
stellar radius can also be indirectly determined from spectra or,
less frequently, directly via interferometry, lunar occultations
or eclipses (e.g. \citealt{fracassini01}). Precisions on stellar
radius measurements of $\sim$3 per cent are now regularly quoted
(e.g. \citealt{fischer05}).

Determining the stellar rotation period,
on the other-hand, tends to be more troublesome. For some active stars,
the stellar spin period can be determined photometrically to high precision
by tracking the passage of large star spots on their surfaces. 
For those systems which do not have photometrically
measured rotation periods, measurements of Ca {\sc ii} H and K emission
can be used to estimate the rotation period by applying the
chromospheric emission -- rotation period relationship
of \cite{noyes84}. Naturally, this latter method is less precise,
and is also affected by intrinsic variability of the Ca {\sc ii} H
and K emission due to, for example, solar-like activity cycles
or the rotation of magnetic regions.

We have carried out an extensive literature search and present $v \sin
i$, $R_*$, and $P_{rot}$ estimates for a number of main-sequence stars
which host spatially resolved debris disks in Table~\ref{tab:1}.
Since one of the pre-requisites for measuring a
stellar rotation period is that the star must be magnetically active,
we are restricted to lower main-sequence stars later than $\sim$F5V
which have a convective envelope (and are thereby capable of
sustaining a stellar dynamo). Of the 20 main-sequence stars with
resolved debris disks, only 10 have spectral types of F5V or later.
Of these, we can find no recorded Ca {\sc ii} H and K emission
measurement for HD 181327, and is therefore omitted from our list.

We should note that we have not considered pre-main sequence stars in
our analysis. This is for two principal reasons. First, given their
fully convective nature, it is not certain that the activity-rotation
period relationship of \cite{noyes84} (which was calibrated for
main-sequence stars) holds, indeed an entirely different stellar
dynamo mechanism may operate in pre-main sequence stars
(e.g. \citealt{scholz07}).
Second, radius estimates for pre-main sequence stars are also
notoriously unreliable, since they depend upon age estimates which
are uncertain by a factor of several (e.g. \citealt{naylor09};
\citealt*{baraffe09}).

\subsection{Adopted stellar parameters and errors}
\label{sec:pars}

In order to determine $\sin i$ via equation~\ref{eqn:sini}, we have taken
a weighted mean of the entries in Table~\ref{tab:1} for the
final values of $v \sin i$ and $R_*$. In identical fashion to that
carried out in \cite{watson10}, where no error was quoted on a 
published $v \sin i$ value we have taken it to be 1.0 kms$^{-1}$ (twice the
typical error assumed on $v \sin i$ measurements, see the catalogue
of \citealt{fischer05} for example). Regarding published radii with no
associated error bar, we have taken the error to be 10 or 20 per cent of the
absolute value. The choice between 10 or 20 per cent is taken to ensure
that radii estimates with associated error bars were given a higher
weighting than those without formal errors.

For stars with photometrically derived rotation periods which have no
associated error bar, we have taken the error to be 10 per cent. This is
commensurate with the typical error bars quoted on such measurements.
Where available, photometrically derived rotation periods are adopted,
otherwise the rotation period is estimated from the strength of the 
Ca {\sc ii} H and K emission (\citealt{noyes84}). Again, following
\cite{watson10}, for each $\log R'_{HK}$ measurement reported in
Table~\ref{tab:1} we have determined, where possible, the number of
observations and period over which they were carried out (see
Table~\ref{tab:rhk}). Where details are not present, or are ambiguous,
we have assumed they are from a single observation and have flagged
them as {\em `individual?'}. As in \cite{watson10}, each star was assigned
a grade of P (Poor), O (O.K.), G (Good) or E (Excellent) based on how well
monitored it was. We then assigned general error bars on the $\log R'_{HK}$
values dependent on their assigned grades and spectral type. These
error bars are derived from the {\em average rotationally modulated
variations} outlined in Section 3.1 of \cite{watson10}. For a detailed
discussion of the systematic errors on the derived parameters, we refer
the reader to this work. Table~\ref{tab:mcmc} lists the adopted parameters
for each star.

\subsection{Determining the stellar inclination angle}

Equation~\ref{eqn:sini} can be thought of as a naive estimator of
$\sin i$ as it is geometrically unconstrained (e.g. $\sin i >$ 1 is
allowed). While a value of $\sin i >$ 1 is unphysical, it does allow
potential problem cases to be identified. Again, we
follow \cite{watson10} and reject systems with $\sin i$'s that are
1-$\sigma$ greater than 1 from further analysis -- flagging these as
having a high probability of being affected by systematic errors.
This results in the omission of 2 systems, HD 53143 and HD 139664, 
both of which have naive $\sin i$ estimates
significantly greater than 1 (see the first two entries of
Table~\ref{tab:mcmc}). In the case of HD 139664, the $B-V$
value places it at the extreme edge of the chromospheric emission --
rotation period calibration by \cite{noyes84}. In addition, the star
is classified as having a luminosity class IV, and therefore both the
derived rotation period from the \cite{noyes84} relationship
(which is only calibrated for main-sequence stars) and radius
may also be suspect. HD 53143, on the other hand, is more problematic.
It appears to have a secure rotation period which has been
measured photometrically and that also agrees very well with the
period derived from the Ca {\sc ii} H and K emission. In addition, all of
the measured $v \sin i$'s and radii are consistent with one another. Yet,
despite this and the fact that it appears to be a solid main-sequence star
with an age of 1.0 $\pm$ 0.2 Gyr (\citealt{kalas06}), we derive $\sin i$ $\sim$
1.5 $\pm$ 0.4. We can only assume that 1 or more of the measurements
are affected by systematics.


For the 8 remaining systems we have carried out a Markov-chain Monte Carlo
(MCMC) analysis which not only provides a means of optimising the fit of a model
to data but explores the joint posterior probability distribution
of the fitted parameters and allows proper 1-$\sigma$ two-tailed
confidence limits to be placed on the derived $\sin i$'s. In addition,
MCMC rejects unphysical combinations of parameters that result in
$\sin i >$ 1. For the purposes of this work, we have followed the MCMC
process outlined in \cite{watson10}, keeping the same 1000-step
burn-in phase and carrying out 1,000,000 jumps. The results of this MCMC
analysis are shown in Table~\ref{tab:mcmc}.

\section{Results and discussion}

Table~\ref{tab:comparison} shows our derived stellar rotation
inclination angles versus published debris-disk inclinations. As can be
seen, there is no obvious evidence for large mis-alignments of the
stellar rotation axes and debris-disk planes in any of these systems.
By the nature of the method, the best constrained systems have
$\sin i \sim$ 0.5 ($i_*$ =30$^{\circ}$). This is because at high
inclinations the sine curve is relatively flat, and thus small
errors in $\sin i$ (which is what is directly calculated from
the observables in equation~\ref{eqn:sini}) propagate to form
large errors when expressed in degrees. At $\sin i$'s of $\sim$0.5,
the sine curve is much steeper, and travelling along the sine
curve does not vary the inclination $i_*$ as quickly as it does at high
$\sin i$'s. As one moves to lower $\sin i$'s, measurement errors on
$v \sin i$ naturally increase as the projected rotational broadening
decreases. The fact that the best constrained systems, HD 22049 and
HD 107146, with errors on $i_*$ of only 5 -- 9$^{\circ}$ appear to align
closely with their debris disk gives us both confidence in the technique, and 
further strengthens our assertion that we see no evidence for a detectable
difference between the sky-projected angle of the disk and the that of the
stellar rotation axis. In addition, it should be noted that HD 22049
is known to host a planet that has had the inclination of its
orbital plane accurately determined to be $i_{planet}$ =
30.$^{\circ}$1$\pm$3.$^{\circ}$8 -- suggesting coplanarity between the planetary
orbit and disk (\citealt{benedict06}). Furthermore, star spot
modeling of a MOST light curve of HD 22049 by \cite{bryce06} determined
the inclination of the stellar rotation axis to be $i_*$ = 
30$^{\circ}\pm$3$^{\circ}$, in excellent agreement with our derived values.
We do caution, however, that the absolute direction of the axis
(whether the rotation axis is pointing towards or away from the observer)
cannot be ascertained, and therefore we do not have a
knowledge of the full three-dimensional geometry of the star-disk systems.

We can test the significance of our result using a rank-order
approach. If the disk and stellar spin axes are closely aligned, we
expect a ranking by disk inclination to agree well with a ranking by
stellar inclination. This is indeed what we find. We calculate the
spearman rank-order correlation coefficient for our data, and find a
value of 0.82. If we repeat this analysis for all 40320 random
permutations of our stellar spin data, only in 0.1\% of cases do we
find a better correlation between stellar spin and disk inclinations.
This suggests we reject the null hypothesis (that stellar spin and disk
inclinations are uncorrelated) at a significance of 99.9\%.

We performed a further test of significance by first assuming that the
stellar spin and disk axes were uniformly distributed around the sky.
We drew 8 values of disk and stellar inclinations from a uniform
distribution in $\cos i$, and took account of our errors by perturbing
these inclinations by an amount equal to the errors on our
observations. For the inclinations, which have asymmetric errors, we
used an average of the two errors. For the disks we assigned an error
of 3 degrees where none was quoted. Where the given disk inclination
is a range, we set the error to half of that range. We computed
$1\times10^6$ such simulated datasets, and only 0.4\% showed a better
correlation (as judged by the Spearman rank-order coefficient) than
our original data. We therefore reject the hypothesis that the stellar
spin axes are independent of the disk inclination axes with a
significance of 99.6\%.

In reality, it is likely that the stellar spin axes and disk axes are
aligned to some degree of precision. It is reasonable to ask what is
the largest degree of misalignment permitted by our data. Answering
such a question would require a full Bayesian treatment which accounts
for the fact that we only observe the component of misalignment along
the line of sight. Such an analysis is quite complex \cite[see][for
example]{fabrycky09}, and is left for a future work.

A recent analysis of Rossiter-McLaughlin observations by
\cite{triaud10} suggest that between 45 -- 85 per cent of hot-Jupiters
appear to be significantly misaligned. Our work in this paper
reveals no similar degree of misalignment between debris disks and their host
stars. We note, however, that all of the systems in our study have
host stars with effective temperatures below $\sim$6140K
(see Table~\ref{tab:comparison}). Recently,
\cite{winn10} have highlighted that exoplanet host stars with
effective temperatures below $\sim$6250K appear to have the planet-star
spin axes preferentially aligned, whereas exoplanets orbiting
hotter host stars are more likely to be misaligned. It is, therefore,
possible that a yet to be determined mechanism which only drives
star-disk misalignments in hotter systems could be operating which
our small study has missed. We conclude, however, that there appears
to be no substantial evidence to suggest that a universal process,
such as that outlined by \cite{Lai2010} and \cite{bate10}, is a major
mechanism in misaligning planetary orbits.

\onecolumn

\begin{center}

\begin{longtable}{lllr@{.}lr@{.}lr@{.}lr@{.}lr@{.}lr@{.}lr@{.}l}
\caption[]{Published data on the properties of 10 stars hosting
  resolved debris disks. Rotation periods quoted with no
reference have been calculated using the adjacent
$\log R'_{HK}$ entry using the \cite{noyes84}
chromospheric emission -- rotation period relationship
along with ($B-V$) values taken from NStED.} \\


\hline  & & \multicolumn{1}{c}{Alternative} & \multicolumn{2}{c}{$v
  \sin i$} & \multicolumn{2}{c}{$\sigma_v$} & \multicolumn{2}{c}{$\log
  R'_{HK}$} & \multicolumn{2}{c}{$P_{rot}$} &
\multicolumn{2}{c}{$\sigma_p$} & \multicolumn{2}{c}{Radius} &
\multicolumn{2}{c}{$\sigma_r$} \\ \multicolumn{1}{c}{HD} &
\multicolumn{1}{c}{HIP} & \multicolumn{1}{c}{Name} &
\multicolumn{2}{c}{(km s$^{-1}$)} & \multicolumn{2}{c}{} &
\multicolumn{2}{c}{} & \multicolumn{2}{c}{(days)} &
\multicolumn{2}{c}{} & \multicolumn{2}{c}{($R_{\odot}$)}
\\ \multicolumn{1}{c}{(1)} & \multicolumn{1}{c}{(2)} &
\multicolumn{1}{c}{(3)} & \multicolumn{2}{c}{(4)} &
\multicolumn{2}{c}{(5)} & \multicolumn{2}{c}{(6)} &
\multicolumn{2}{c}{(7)} & \multicolumn{2}{c}{(8)} &
\multicolumn{2}{c}{(9)} & \multicolumn{2}{c}{(10)}  \\ \hline
\endfirsthead

\multicolumn{17}{c} {{\bfseries \tablename\ \thetable{} -- continued}}
\\ \hline

& & \multicolumn{1}{c}{Alternative} & \multicolumn{2}{c}{$v \sin i$} &
\multicolumn{2}{c}{$\sigma_v$} & \multicolumn{2}{c}{$\log R'_{HK}$} &
\multicolumn{2}{c}{$P_{rot}$} & \multicolumn{2}{c}{$\sigma_p$} &
\multicolumn{2}{c}{Radius} & \multicolumn{2}{c}{$\sigma_r$} \\

\multicolumn{1}{c}{HD} & \multicolumn{1}{c}{HIP} &
\multicolumn{1}{c}{Name} & \multicolumn{2}{c}{(km s$^{-1}$)} &
\multicolumn{2}{c}{} & \multicolumn{2}{c}{} &
\multicolumn{2}{c}{(days)} & \multicolumn{2}{c}{} &
\multicolumn{2}{c}{($R_{\odot}$)} \\

\multicolumn{1}{c}{(1)} & \multicolumn{1}{c}{(2)} &
\multicolumn{1}{c}{(3)} & \multicolumn{2}{c}{(4)} &
\multicolumn{2}{c}{(5)} & \multicolumn{2}{c}{(6)} &
\multicolumn{2}{c}{(7)} & \multicolumn{2}{c}{(8)} &
\multicolumn{2}{c}{(9)} & \multicolumn{2}{c}{(10)} \\ \hline \endhead

10647.....&7978....&&5&600$^2$&0&500&-4&68$^{8}$&7&562&.&.&1&080$^1$&0&050 \\
&&&6&000$^4$&.&.&-4&700$^{9}$&7&903&.&.&0&990$^{12}$&.&. \\
&&&4&880$^{8}$&.&.&-4&714$^{11}$&8&137&.&.&1&096$^{13}$&0&025 \\
&&&5&200$^{11}$&.&.&.&.&.&.&.&.&1&14$^{14}$&0&040 \\

10700.....&8102....&TAU Cet&1&300$^2$&0&500&-4&980$^3$&32&848&.&.&0&750$^1$&0&030 \\
&&&1&000$^4$&.&.&-4&955$^5$&32&058&.&.&0&880$^{17}$&0&100 \\
&&&0&800$^{7,a}$&0&400&-4&958$^{22}$&34&00$^{22,p}$&.&.&0&830$^{14}$&0&020 \\
&&&2&000$^{24}$&.&.&-4&955$^{23}$&32&058&.&.&.&.&.&. \\
&&&0&400$^{25}$&0&400&-5&026$^{11}$&34&266&.&.&.&.&.&. \\

22049.....&16537...&Epsilon Eri&2&400:$^2$&.&.&-4&510:$^3$&17&275:&.&.&0&740:$^1$&0&030 \\
&&&1&700$^{25}$&0&300&-4&455$^{22}$&12&000$^{22,p}$&.&.&0&860$^{17}$&0&120 \\
&&&1&800$^{7,a}$&0&400&.&.&11&300$^{26,p}$&1&100&0&690$^{12}$&.&. \\
&&&.&.&.&.&.&.&11&150$^{27,p}$&1&150&0&770$^{14}$&0&020 \\
&&&.&.&.&.&.&.&11&300$^{23,p}$&.&.&.&.&.&. \\

53143.....&33690...&&4&000$^4$&.&.&-4&520$^5$&16&298&.&.&0&920$^1$&0&050 \\
&&&4&100$^{11}$&.&.&-4&507$^{11}$&15&528&.&.&0&880$^{12}$&.&. \\
&&&4&000$^{10}$&.&.&.&.&16&400$^{18,p}$&.&.&0&870$^{17}$&.&. \\
&&&.&.&.&.&.&.&.&.&.&.&0&850$^{13}$&0&020 \\

61005.....&36948...&&9&000$^4$&.&.&-4&260$^5$&3&677&.&.&0&810$^{12}$&.&. \\
&&&8&200$^{11}$&.&.&-4&324$^{11}$&5&551&.&.&0&840$^{1}$&0&06 \\
&&&.&.&.&.&-4&360$^{15}$&6&826&.&.&.&.&.&. \\
&&&.&.&.&.&-4&337$^{16}$&5&993&.&.&.&.&.&. \\

92945.....&52462...&GJ 3615&4&000$^4$&.&.&-4&320$^3$&6&964&.&.&0&810$^1$&0&050 \\
&&&5&100$^2$&0&500&-4&393$^{16}$&10&446&.&.&0&780$^{14}$&0&030 \\
&&&5&100$^{7,a}$&2&100&.&.&13&470:$^{21}$&.&.&0&770$^{12}$&.&. \\
&&&4&000$^{10}$&.&.&.&.&.&.&.&.&.&.&.&.\\

107146..&60074..&&5&000$^2$&0&500&-4&340$^3$&3&496&.&.&0&990$^1$&0&070 \\
&&&5&000$^4$&.&.&.&.&.&.&.&.&0&981$^{2}$&0&027 \\
&&&.&.&.&.&.&.&.&.&.&.&1&000$^{14}$&0&040 \\
&&&.&.&.&.&.&.&.&.&.&.&1&000$^{13}$&0&020 \\
&&&.&.&.&.&.&.&.&.&.&.&0&970$^{12}$&.&. \\

139664..&76829.....&GJ 594&71&600$^6$&3&600&-4&621$^{11}$&1&517&.&.&1&33$^1$&0&060 \\
&&&105&000$^7$&.&.&.&.&.&.&.&.&1&270$^{17}$&0&500 \\
&&&87&000$^{19}$&.&.&.&.&.&.&.&.&1&318$^{13}$&0&030 \\
&&&.&.&.&.&.&.&.&.&.&.&1&260$^{12}$&.&. \\

197481....&102409..&AU Mic&9&300$^{10}$&1&2&-4&520$^5$&4&865$^{21,p}$&.&.&0&870$^1$&0&020 \\
&&&8&000:$^{7}$&.&.&.&.&4&850$^{18,p}$&.&.&0&860$^{12}$&.&. \\
&&&.&.&.&.&.&.&4&822218$^{20,p}$&.&.&0&610$^{17}$&0&050 \\
207129 & 107649 & GJ 838 & 2&000$^4$ & 0&000 &-4&800$^5$&15&171&.&.&1&040$^1$&0&050 \\
&&&2&400$^2$&0&500&-4&850$^9$&16&296&.&.&0&985$^{17}$&.&.\\
&&&.&.&.&.&-5&020$^{16}$&19&536&.&.&1&080$^{14}$&0&040 \\
&&&.&.&.&.&.&.&.&.&.&.&1&047$^{13}$&0&024 \\
&&&.&.&.&.&.&.&.&.&.&.&0&980$^{12}$&.&. \\
\label{tab:1}

\end{longtable}
\end{center}

\vspace{-1.1cm}

\noindent References: $^1$NStED, $^2$\cite{valenti05}, $^3$\cite{wright04},
$^4$\cite{nordstroem04}, $^5$\cite{henry96}, $^6$\cite{reiners03b},
$^{7}$\cite{glebocki00}, $^{8}$ Coralie,
$^{9}$\cite{jenkins06}, $^{10}$\cite{torres06},
$^{11}$\cite*{schroder09}, $^{12}$\cite{rhee07},
$^{13}$\cite{allende99}, $^{14}$\cite{takeda07},
$^{15}$\cite{white07}, $^{16}$\cite{gray06},
$^{17}$\cite{fracassini01}, $^{18}$\cite{pizzolato03},
$^{19}$\cite{ochsenbein99}, $^{20}$\cite{pojmanski05},
$^{21}$\cite{samus09}, $^{22}$\cite{baliunas96},
$^{23}$\cite{noyes84}, $^{24}$\cite{mallik03}, $^{25}$\cite{saar97},
$^{26}$\cite{simpson10a}, $^{27}$\cite{fray91}

\noindent  : = value uncertain. $^a$ = mean of a range of values
given by \cite{glebocki00}. $^p$ = rotation period measured photometrically.

\twocolumn

\onecolumn

\begin{center}

\begin{longtable}{lr@{.}lr@{.}lr@{.}lr@{.}lr@{.}lr@{.}lr@{.}lr@{.}lr@{.}lr@{.}l}
\caption[]{Adopted parameters and $\sin i$ estimates for the stars in
  our study.  The colour index $(B-V)$ is only indicated for stars
  where the rotation period has been determined from the chromospheric
  activity -- rotation period relationship of \cite{noyes84}. The
  first two entries have `naive' $\sin i$ estimates in column 9
  (indicated by an asterisk) as derived from equation~\ref{eqn:sini}
  which, complete with their formally propagated errors, result in
  $\sin i$'s significantly above 1. These stars are omitted from
  further analysis. For the remaining eight stars, column 9 gives the
  final derived $\sin i$ value, followed by the 1-$\sigma$ two-tailed
  confidence limits, as derived from a Markov-chain Monte Carlo
  analysis.} \\

\hline \multicolumn{1}{c}{HD or} & \multicolumn{2}{c}{$v \sin i$} &
\multicolumn{2}{c}{$\sigma_v$} & \multicolumn{2}{c}{$P_{rot}$} &
\multicolumn{2}{c}{$\sigma_P$} & \multicolumn{2}{c}{$R_*$} &
\multicolumn{2}{c}{$\sigma_R$} & \multicolumn{2}{c}{$B-V$} &
\multicolumn{2}{c}{$\sin i$} & \multicolumn{2}{c}{$\sigma_-$} &
\multicolumn{2}{c}{$\sigma_+$} \\

\multicolumn{1}{c}{Alt. Name} & \multicolumn{2}{c}{(km s$^{-1}$)} &
\multicolumn{2}{c}{}& \multicolumn{2}{c}{(days)} &
\multicolumn{2}{c}{} & \multicolumn{2}{c}{($R_{\odot}$)} &
\multicolumn{2}{c}{} & \multicolumn{2}{c}{} & \multicolumn{2}{c}{} &
\multicolumn{2}{c}{} & \multicolumn{2}{c}{}\\

\multicolumn{1}{c}{(1)} & \multicolumn{2}{c}{(2)} &
\multicolumn{2}{c}{(3)}& \multicolumn{2}{c}{(4)} &
\multicolumn{2}{c}{(5)} & \multicolumn{2}{c}{(6)} &
\multicolumn{2}{c}{(7)} & \multicolumn{2}{c}{(8)} &
\multicolumn{2}{c}{(9)} & \multicolumn{2}{c}{(10)} &
\multicolumn{2}{c}{(11)} \\

\hline

53143       &   4 & 033 & 1 & 000 &  16 & 399 &  1 & 639 &  0 & 850 & 0 & 019 & .&. & 1 & 536$^*$ & 0 & 412 & 0 & 412 \\ 
139664      &  89 & 711 & 1 & 827 &   1 & 517 &  0 & 249 &  1 & 319 & 0 & 026 & 0&400 & 2 & 038$^*$ & 0 & 339 & 0 & 339 \\
10647       &   5 & 497 & 0 & 377 &   7 & 803 &  1 &  32 &  1 & 099 & 0 & 019 & 0&534 & 0 & 768 & 0 & 142 & 0 & 157 \\
10700       &   0 & 848 & 0 & 232 &  34 & 000 &  3 & 399 &  0 & 807 & 0 & 016 & .&. & 0 & 702 & 0 & 208 & 0 & 229 \\
22049       &   1 & 772 & 0 & 233 &  11 & 300 &  0 & 510 &  0 & 770 & 0 & 019 & .&. & 0 & 510 & 0 & 071 & 0 & 081 \\
61005       &   8 & 599 & 1 & 000 &   5 & 419 &  2 & 108 &  0 & 829 & 0 & 048 & 0&742 & 0 & 999 & 0 & 123 & 0 & 000 \\
92945       &   5 & 022 & 0 & 468 &   7 & 176 &  2 & 830 &  0 & 786 & 0 & 024 & 0&894 & 0 & 908 & 0 & 091 & 0 & 087 \\
107146      &   5 & 000 & 0 & 447 &   3 & 496 &  1 &  35 &  0 & 993 & 0 & 014 & 0&611 & 0 & 353 & 0 & 141 & 0 & 138 \\
197481      &   8 & 832 & 0 & 959 &   4 & 846 &  0 &  20 &  0 & 835 & 0 & 018 & .&. & 0 & 999 & 0 & 062 & 0 & 000 \\
207129      &   2 & 319 & 0 & 447 &  17 & 129 &  1 & 610 &  1 & 048 & 0 & 018 & 0&600 & 0 & 746 & 0 & 167 & 0 & 187 \\

\label{tab:mcmc}

\end{longtable}

\end{center}

\twocolumn

\begin{table}
\caption[]{Comparison of the derived stellar rotational axes and published
disk-plane inclinations. For HD 10647 and HD 10700 the lower value for the
disk inclination corresponds to that derived from the observed disk
dimensions and which we take to be the most probable value. References
for the disk inclinations are given in the fourth column. Estimates of the
host star mass and effective temperature from the NStED database are quoted in
the final two columns.}
\begin{tabular}{llcccc}
\hline
HD & $i_*$ ($^{\circ}$) & $i_{disk}$ ($^{\circ}$) & ref. & Mass ($M_{\odot}$) & T$_{eff} (K)$\\

\hline
10647 & 49$^{+17}_{-11}$ & $\geq$52 & (\citealt{liseau08}) & 1.20 & 6140 \\
10700 & 45$^{+24}_{-15}$ & 60--90 & (\citealt{greaves04}) & 0.91 & 5500 \\
22049 & 31$^{+5}_{-5}$ & 25 & (\citealt{greaves98}) & 0.78 & 5090 \\
61005 & 90$^{+0}_{-26}$ & 80 & (\citealt{maness09}) & 0.89 & 5440 \\
92945 & 65$^{+21}_{-10}$ & 70 &  (\citealt{krist05}) & 0.77 & 5060 \\
107146 & 21$^{+8}_{-9}$ & 25$\pm$5 & (\citealt{ardila04}) & 1.09 & 5850 \\
197481 & 90$^{+0}_{-20}$ & 90 &  (\citealt{krist05b}) & 0.49 & 3560 \\
207129 & 47$^{+22}_{-13}$  & 60$\pm3$ & (\citealt{krist10}) & 1.11 & 5890 \\

\hline

\label{tab:comparison}

\end{tabular}
\end{table}

\onecolumn

\begin{center}
\begin{longtable}{lccll}

\caption[]{Compilation of chromospheric indices ($\log R'_{HK}$) for
  the stars in Table 1 for which no measured rotation periods have been
  reported. The spectral type of the host star is given in
  column 2. Entries in bold give the grade assigned to each star (P =
  Poor, O = O.K., G = Good, and E = Excellent) followed by the
  weighted mean of the $\log R'_{HK}$ measurements and adopted error
  bar (see section~\ref{sec:pars} for details). Reference numbers
  are identical to those used in Table~\ref{tab:1}.}

\\

\hline 
\hline
Name & Type & $\log$ R'$_{HK}$ & Observations & Ref. \\
\hline
\endfirsthead

\multicolumn{5}{c} {{\bfseries \tablename\ \thetable{} -- continued}}
\\ \hline
\hline
Name & Type & $\log R'_{HK}$ & Observations & Ref.  \\
\hline \endhead

HD 10647 & F8V & -4.680 & individual? & 8  \\
         &     & -4.700 & individual on 2001 Aug 04 & 9. \\
         &     & -4.714 & individual? & 11. \\
& & & {\bf (P) Adopted value: -4.698 $\pm$ 0.060} \\
\hline
HD 61005 & G3/5V & -4.260 & 1 obs on UT 14/12/1992 & 5 \\
& & -4.324 & individual? & 11 \\
& & -4.360 & 1 obs on 28/10/2002 & 15 \\
& & -4.337 & individual? & 16 \\
& & & {\bf (P) Adopted value: -4.320 $\pm$ 0.075} \\
\hline
HD 92945 & K1V & -4.320 & 13 obs in 6 months. Report $\sigma$ = 2.72\% & 3 \\
& & -4.393 & individual? & 16 \\
& & & {\bf (O) Adopted value: -4.325 $\pm$ 0.077} \\
\hline
HD 107146 & G5 & -4.340 & 8 obs in 5 months. Report $\sigma$ = 3.04\% & 3 \\
& & & {\bf (O) Adopted value: -4.340 $\pm$ 0.057} \\
\hline
HD 139664 & F3/5V & -4.621 & individual? & 11 \\
& & & {\bf (P) Adopted value: -4.621 $\pm$ 0.060} \\
\hline
HD 207129 & G0V & -4.800 & 1 obs on UT 28/06/1993 & 5 \\
& & -4.850 & 1 obs on 2004 Aug 23/24 & 9 \\
& & -5.020 & individual? & 16 \\
& & & {\bf (P) Adopted value: -4.89 $\pm$ 0.075} \\
\hline

\label{tab:rhk}

\end{longtable}

\end{center}

\twocolumn

\section*{\sc Acknowledgements}

CAW and CD would like to thank the Nuffield Foundation for funding
a science bursary for CD to undertake a summer placement at QUB.
SPL acknowledges the support of an RCUK fellowship. Finally, we
would like to thank Dan Fabrycky for his comments which helped to
substantially improve the conclusions of this paper.

\bibliographystyle{mn2e}
\bibliography{abbrev,refs}

\end{document}